\documentclass[aps,prb,a4paper,twocolumn,floatfix,showkeys,amsmath,amssymb,nofootinbib,nobibnotes,altaffilletter]{revtex4}

\usepackage[T1]{fontenc}
\usepackage[latin1,utf8]{inputenc}
\usepackage[english]{babel}

\usepackage{setspace}

\usepackage{amsmath}
\usepackage{bm}

\usepackage{graphicx}
\usepackage{pslatex}
\usepackage{xspace}
\usepackage{subfigure}
\usepackage{float}

\usepackage{booktabs}

\usepackage{ragged2e}
\usepackage[labelfont=bf, format=plain, indention=0.5cm, textfont=small, skip=4pt, justification=RaggedRight, tableposition=top, figureposition=bottom ]{caption}

\newcommand{\pht}{poly(3-hexyl thiophene)\xspace}
\newcommand{\pcbm}{[6,6]-phenyl-C$_{61}$ butyric acid methyl ester\xspace}

\newcommand{\figref}[1]{Fig.~\ref{fig:#1}}

\renewcommand{\eqref}[1]{Eq.~(\ref{eq:#1})}

\begin{document}

\preprint{}

\title{Role of Polaron Pair Diffusion and Surface Losses in Organic Semiconductor Devices}

\author{Thomas \surname{Strobel}}
\author{Carsten \surname{Deibel}}\email{deibel@physik.uni-wuerzburg.de}
\affiliation{Experimental Physics VI, Julius-Maximilians-University of W{\"u}rzburg, 97074 W{\"u}rzburg, Germany}

\author{Vladimir \surname{Dyakonov}}
\affiliation{Experimental Physics VI, Julius-Maximilians-University of W{\"u}rzburg, 97074 W{\"u}rzburg, Germany}
\affiliation{Bavarian Centre for Applied Energy Research (ZAE Bayern), 97074 W{\"u}rzburg, Germany}

\date{November 19, 2010}


\begin{abstract}
By applying Monte Carlo simulations we found that the extraction of bound polaron pairs (PP) at the electrodes is an important loss factor limiting the efficiency of organic optoelectronic and photovoltaic devices.
Based upon this finding, we developed a unified analytic model consisting of exact Onsager theory,  describing the dissociation of PP in organic donor--acceptor heterojunctions, the Sokel--Hughes model for the extraction of free polarons at the electrodes, as well as of PP diffusion leading to the aforementioned loss mechanism, which was not considered previously.
Our approach allows to describe the simulation details on a macroscopic scale and to gain fundamental insights, which is important in view of developing an optimized photovoltaic device configuration. 
\end{abstract}



\maketitle


In solar cells based on polymer--fullerene bulk heterojunctions (BHJ) primary molecular excitations, also called excitons, are generated by light absorption.
The photogenerated excitons diffuse by emission and re-absorption to donor-acceptor interfaces. \cite{hwang2008}
There, the excitations may undergo charge transfer from the excited donor molecules to the nearby acceptors. \cite{sariciftci1992,deibel2010review}
The resulting charge transfer states are also called polaron pairs (PP) and, unlike bipolarons, consist of oppositely charged constituents. They may dissociate into free polarons which can be extracted from the device to generate a net photocurrent.\cite{deibel2010review2} 
In organic solar cells the net photocurrent is crucially determined by the PP dissociation and extraction yields, currently leading to overall power conversion efficiencies of up to $8\%$. \cite{green2010review}

It is commonly implied that PP either recombine or dissociate at the place of their creation, as their diffusion is not accounted for in theoretical models. \cite{onsager1934,braun1984,wojcik2009} However, it is known from experiments that photogenerated charges have very high initial mobilities, \cite{devizis2009} and there are indications that photogenerated charges can move to different molecules or conjugation segments before they recombine or dissociate.\cite{bakulin2009a}
In this Letter the diffusion of Coulomb bound PP resulting in an important loss mechanism in organic BHJ solar cells is considered. We perform Monte Carlo simulations of PP recombination and photocurrent extraction in BHJ, focussing on PP diffusion and the resulting losses at the electrodes.
We find that without blocking layers, over $40\%$ of the generated PP can get lost at the electrodes and that, for high electric fields, this loss mechanism becomes dominant.
Such high surface loss cannot be explained by the theories commonly used. Our simulations show that PP diffusion is responsible, as it can increase extraction of both (bound) charges at the same electrode.
Upon this finding we present a unified analytic model to calculate the photocurrent yield, accounting for PP diffusion and losses at the surface.


In our Monte Carlo simulations we describe the organic blend of P3HT (\pht) and PCBM (\pcbm), a widely studied representative of polymer--fullerene semiconductors, on a mesoscopic scale. \cite{deibel2009a}
Therefore, electron donor and acceptor molecules were distributed in a volume ratio of $1:1$ within a $100\times 25\times 25$ cubic lattice with constant spacing of $1$~nm, sandwiched between electrodes on the long lattice side and periodic boundary conditions for the other directions.
Each acceptor molecule was assigned to a single cubic lattice site, in accordance with  the spherical shape of fullerenes.
However, as conjugated polymers consist of several monomer units and are known to have an effective conjugation length (CL), a donor molecule was spanned over multiple lattice points. \cite{deibel2009a}

Charge transport was described by the Miller--Abrahams\cite{miller1960,bassler1993} rate equation, also accounting for charge carrier Coulomb interaction as well as mirror charge effects. Miller--Abrahams theory was used as it neglects all polaronic effects and thus qualitatively better accounts for PP excess energy, which is initially (over-)compensating molecular reorganization energies in e.g. classical Marcus theory.
In order to include energetic disorder, the energy levels, corresponding to the respective molecular orbitals, were taken from Gaussian distribution functions according to our experimental findings, with standard deviation $\sigma_A=60$~meV for acceptor molecules and $\sigma_D = 75$~meV for donors. \cite{deibel2009a} 
The PP lifetime was set to $\tau_{eff} = 10^{-7}$~s. \cite{mihailetchi2004,veldman2008,deibel2009a}
At simulation initialization, one polaron pair was set at a randomly chosen donor--acceptor interface, which corresponds to a PP density of $n_{PP}=1.6\cdot10^{16}$~cm$^{-3}$.
To gain statistically reliable results, about $2.5\cdot10^{5}$ simulations were performed for each parameter set.

\begin{figure}[tb]
	\includegraphics[width=6.8cm]{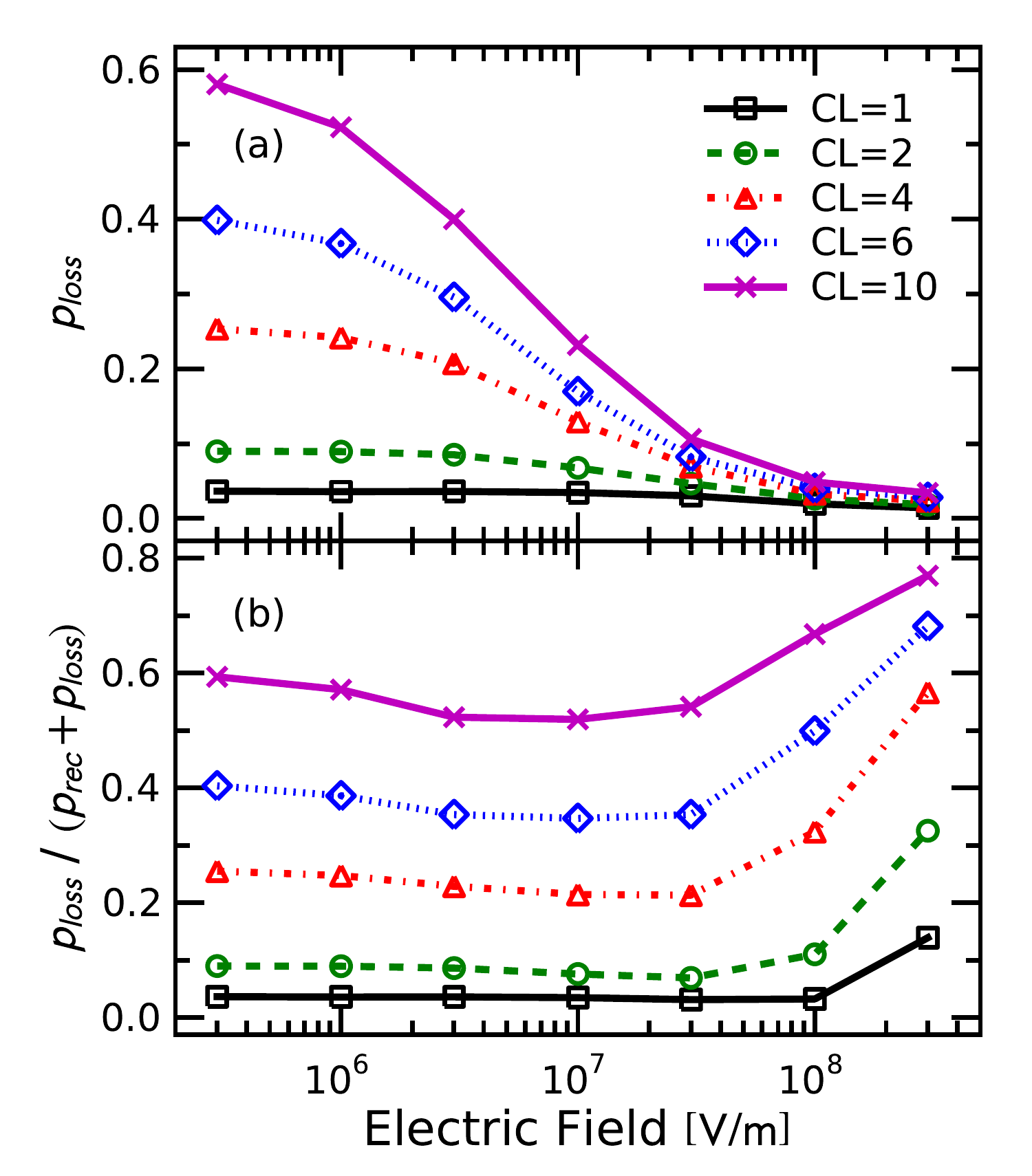}
	\caption{(Color Online) Probability of PP surface losses vs.\ electric field found by Monte Carlo simulation of $1:1$ donor--acceptor blends at $300$~K with $\tau_{eff}=100$~ns (device length $l=100$~nm, dielectric constant $\epsilon=3.0$). (a) Absolute probability for PP surface loss $p_{loss}$ is significant at low fields and for long donor conjugation lengths (CL). (b) $p_{loss}$ relative to the sum of all losses---in the bulk ($p_{rec}$) and at the surface---becomes the dominant loss mechanism for increasing CL and increasing field.\label{fig:1}}
\end{figure}

A crucial step during the photocurrent generation is the PP dissociation yield, $p_{int}$.
We point out that $p_{int}$ corresponds to internal polaron pair dissociation, where surface effects and losses are neglected. We considered it in our former work, \cite{deibel2009a} in which charge carrier delocalisation, implying high local mobilities, was found to be a key parameter for achieving high $p_{int}$. In this Letter, we consider losses at the electrodes, and therefore define the charge extraction yield, $p_{ext}$, which is a measure for the photocurrent. We show that by increasing CL, the probability for losing PP at the electrodes, $p_{loss} = p_{int}-p_{ext}$, increases (\figref{1}a).
Even more, the fraction of PP losses at the surface $p_{loss}$ becomes dominant for CL=10 over all electric fields, as compared to the sum of all bulk ($p_{rec} = 1 - p_{int}$) and surface PP loss mechanism.

\begin{figure}[tb]
	\includegraphics[width=6.8cm]{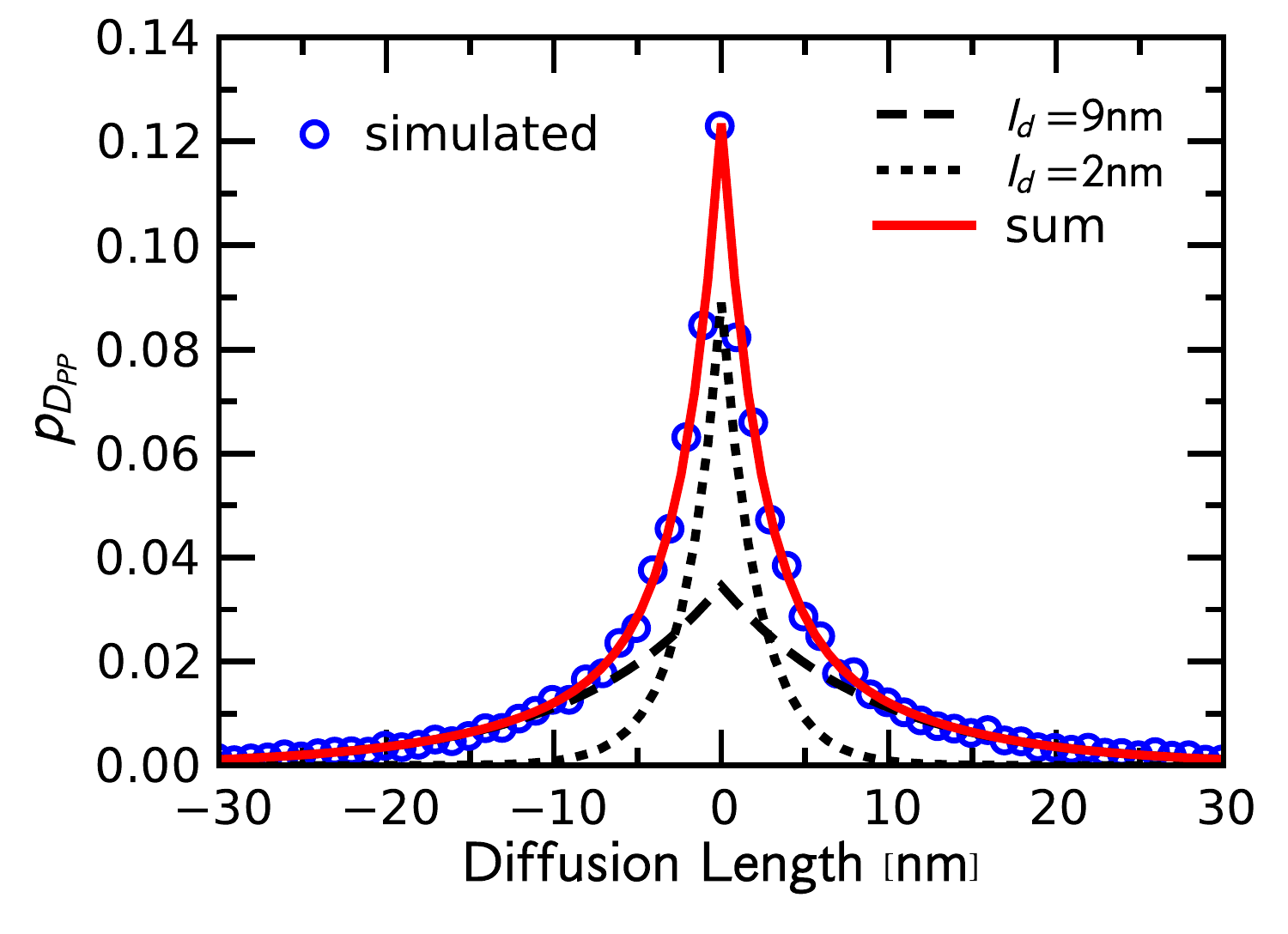}
	\caption{(Color Online) Probability distribution $p_{D_{PP}}$ of polaron pair diffusion lengths from Monte Carlo simulation (circles) of a $1:1$ donor--acceptor blend at $300$~K with $\tau_{eff}=100$~ns (device length $l=100$~nm, dielectric constant $\epsilon=3.0$). The dashed lines indicate the fitted contributions of the two identified diffusion processes, with diffusion lengths of $2$~resp.\ $9$~nm, the solid line representing the sum of both.\label{fig:2}}
\end{figure}

Investigating this significant loss mechanism in detail showed that the PPs do not necessarily dissociate or recombine at the place of their creation.
Due to their strong mutual Coulomb binding, the generated polaron pairs diffuse as neutral quasiparticles within their lifetimes along the distributed donor--acceptor interfaces.
To illustrate this effect, the diffusion of external field stabilized PPs in direction of that external field of $3\cdot10^5$~V/m is shown in \figref{2} exemplarily for $\text{CL}=6$. The graph shows the fraction of PP vs.\ their diffusion length before recombination occurs. Only a tenth of the PP does not diffuse, as indicated by the peak of the probability distribution.
For longer CL, the PP diffusion length increases, whereas higher external fields reduce the PP bulk recombination in general (not shown).
Thus, the surface loss of PP increases at the same time as bulk recombination $p_{rec}$ reduces with increasing CL. \cite{deibel2009a}

As the diffusion lengths distribution was averaged over more than $10^5$ simulations, the discrete random-walk diffusion process of the simulation can be approximated by a continuous-time stochastic Wiener process, also known as Brownian motion.
For PP, decaying exponentially with $\tau_{eff}$, the probability to diffuse for the distance $x$ is given as
\begin{equation}\label{eq:diffusion}
	p_{D_{PP}}( x ) =  \int\limits_0^{\infty} \exp \left( - \frac{ t }{ \tau_{eff} } \right) \cdot \frac{1}{\sqrt{2\pi \cdot 2 D_{PP} t } } \exp \left( - \frac{1}{2} \frac{ {x}^2 }{ 2 D_{PP} t } \right) dt.
\end{equation}
Thereby $D_{PP}$ is the PP diffusion constant, which relates to the characteristic diffusion length $l_d$ of the process as $l_d = \sqrt{ D_{PP} \cdot \tau_{eff} }$ .

We find that the simulated distribution can be expressed well as the combination of two diffusion processes of different $D_{PP}$, as shown in~\figref{2}.
For the given parameters, $40\%$ of the PP hardly diffuse and recombine within a distance of $l_d = 2$~nm.
The other $60\%$ diffuse for $l_d = 9$~nm, in total corresponding to $12\%$ surface loss due to PP diffusion (device length $100$~nm). For thinner devices, the relative losses are even more significant (not shown).
We observe that the two diffusion processes found are related to the energetic relaxation of charge carriers in the initially equally occupied density of acceptor or donor states.

To study these high surface losses found in our simulations in more detail, and to identify the essential contributing processes, we include the PP diffusion into an analytic model. 

\begin{figure}[tb]
	\includegraphics[width=6.8cm]{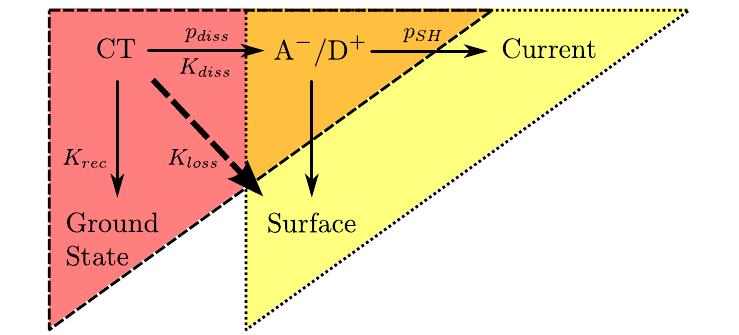}
	\caption{(Color Online) Combination of dissociation theory (left triangle) and drift and diffusion model (right triangle). PP dissociation probability $p_{diss}$ results from competitive rates of the individual PP loss processes $K_i$ and leads to net current with $p_{SH}$. Direct surface loss of PP ($K_{loss}$) has not been addressed before.\label{fig:3}}
\end{figure}

Our approach is shown schematically in~\figref{3} and is as follows: Starting from the created PP, first their change in number is described by competing rates (\figref{3}, left triangle), similar to Onsager--Braun \cite{onsager1934,braun1984} theory. However, by not restricting our approach to time independent rate constants, we can include a time dependent surface loss process for the photogenerated PP. In contrast, Onsager--Braun or exact Onsager \cite{wojcik2009,noolandi1979} extensions are entirely done on infinite space, thus completely neglecting surface losses due to space confinement. \cite{onsager1934,noolandi1979}
Second, the final extraction yield of the individual charge carriers created by PP dissociation is calculated with a drift and diffusion model for the individual charge carriers (\figref{3}, right triangle). \cite{sokel1982,mihailetchi2004}

In our general, time dependent approach, the probability $p_i$ of a PP to decay by a certain process $i$---such as recombination, dissociation or surface loss---is the integral over the relative probability $k_i(t)$ of that individual decay process and the probability $n(t)$ that the PP still exists at the time $t$ after its creation,
\begin{equation}\label{eq:p_i}
	p_i = \int_0^{\infty} k_i(t) n(t) dt \quad .
\end{equation}
The time dependence of $k_i(t)$ is relative to the creation of the PP and not due to external changes on the whole system, thus equal for each PP. Time independent $k_i$ are also often called (decay) rates and---assuming non-interacting PP initially created at $t=0$s---$n(t)$ is called the number of (existing) PP.

The competing, time dependent decay probabilities $k_i(t)$ result in the general differential equation for $n(t)$,
\begin{equation}
	\dot{n}(t) = - \sum_{i} k_i(t) \cdot n(t) \quad ,
	\label{eq:differential_equation}
\end{equation}
where the formal solution for the initial value of $n(0) = 1$ can be written as
\begin{equation}\label{eq:n}
	n(t) = \prod_{i} K_i(t) \quad \text{with} \quad K_i(t) = e^{-k_i(t) \cdot t} \quad .
\end{equation}

We note that for time independent $k_i$ the integral over $n(t)$ in \eqref{p_i} results in the inverse sum of $k_i$ and so yields the common equation of competing rates, as e.g. known from Onsager--Braun theory.

In order to calculate the probabilities of the PP losses $p_i$ according to~\figref{3} (left triangle), $k_i(t)$ or equivalent $K_i(t)$ for recombination, surface loss and dissociation process have to be determined.

Commonly, recombination processes are described according to exponential decay with a recombination rate $k_{rec}$, which is associated with the inverse PP lifetime $\tau$ by $k_{rec} = 1/\tau$.

In order to calculate the PP dissociation rate $k_{diss}$, we start from an exact solution of Onsager theory \cite{noolandi1979} to get an expression for the PP dissociation probability which contains recombination and dissociation losses, but neglects any surface effects.
In detail, the PP dissociation probability $\varphi_{diss}( k_{rec}, E, r_{PP}, \mu )$ accounts for Langevin type recombination, finite recombination rates and finite recombination distances and mainly depends on recombination rate $k_{rec}$, external field $E$, initial PP radius $r_{PP}$ and charge carrier mobility $\mu$. \cite{wojcik2009}
In a second step, the dissociation rate $k_{diss}$ is obtained out of the dissociation probability $\varphi_{diss}$ by inverting its equivalent expression as competing rates of $k_{rec}$ and $k_{diss}$, thus
\begin{equation}\label{eq:k_diss}
	k_{diss} = \frac{ \varphi_{diss} \cdot k_{rec} }{ 1 - \varphi_{diss} } \quad .
\end{equation}
Alternatively, $k_{diss}$ could be approximated directly with Onsager--Braun. \cite{braun1984}

In the newly considered surface loss process a PP is lost when it diffused to one of the electrodes.
As the PP are created uniformly within the device, surface loss is the dominant loss mechanism at early times because PP generated close to the electrodes most likely get lost. However, at longer times surface loss rapidly becomes less significant (not shown), as the PP loss probability changes during the lifetime of the PP.
In order to describe the time dependent surface loss process due to PP diffusion with the diffusion coefficient $D_{PP}$, the same continuous-time stochastic Wiener process as in \eqref{diffusion} is used,
\begin{equation}\label{eq:K_loss}
	K_{loss}(t) =  \frac{1}{l_{eff}}\int\limits_0^{l_{eff}} \int\limits_0^{l_{eff}} \frac{1}{\sqrt{2\pi \cdot 2 D_{PP} t } } \exp \left( - \frac{1}{2} \frac{ {(x - \bar{x})}^2 }{ 2 D_{PP} t } \right) dx d\bar{x} .
\end{equation}
The real device length has to be taken relative to the average hopping distance of the charge carriers, because the Wiener process is the continuous limit of a discrete random walk process. We express this by a virtual device shortening to an effective device length $l_{eff}$.
Thus, the importance of the surface loss process is proportional to the ratio of diffusion length $l_d$ to the effective device length $l_{eff}$.

Now, $p_{diss}$ and $p_{int} = 1 - p_{rec}$ can be calculated by integrating \eqref{p_i} with $k_{rec}$, $k_{diss}$ and $K_{loss}(t)$ from above, which is the last step in calculating the net generation of free charge carriers (\figref{3}, left triangle).
We note that for $k_{loss}(t) \equiv 0$, i.e., $K_{loss}(t) \equiv 1$, \eqref{p_i} simplifies to exact Onsager or Onsager--Braun theory, depending on which theory was used to describe $k_{diss}$.

Once a PP is dissociated into free polarons, these free charge carriers have to be extracted over the electrodes to contribute to the net photocurrent (\figref{3}, right triangle).
Unlike the pure PP diffusion of neutral quasiparticles, the individual free charges are affected by the external field.

A model for non-interacting charge carriers describing charge extraction in a device due to drift and thermal diffusion is that from Sokel and Hughes. \cite{sokel1982}
Assuming a uniform charge carrier generation without recombination losses within the device, the probability $p_{coll}$ of collecting charges with an external electric field $E$ on the corresponding electrodes of a device of real lenght $l$ is 
\begin{equation}
	p_{coll} = \frac{ \exp(eEl/k_\text{B}T) + 1 }{ \exp(eEl/k_\text{B}T) - 1 } - \frac{ 2k_\text{B}T }{eEl} \quad .
\end{equation}
The probability of losing individual, free charges on the wrong electrodes is simply $1 - p_{coll}$ (\figref{3}, right triangle).

The total probability of net photocurrent generation is finally given as the product of net generation and collection probabilities as $p_{ext} = p_{coll} \cdot p_{diss}$ .

\begin{figure}[tb]
	\includegraphics[width=6.8cm]{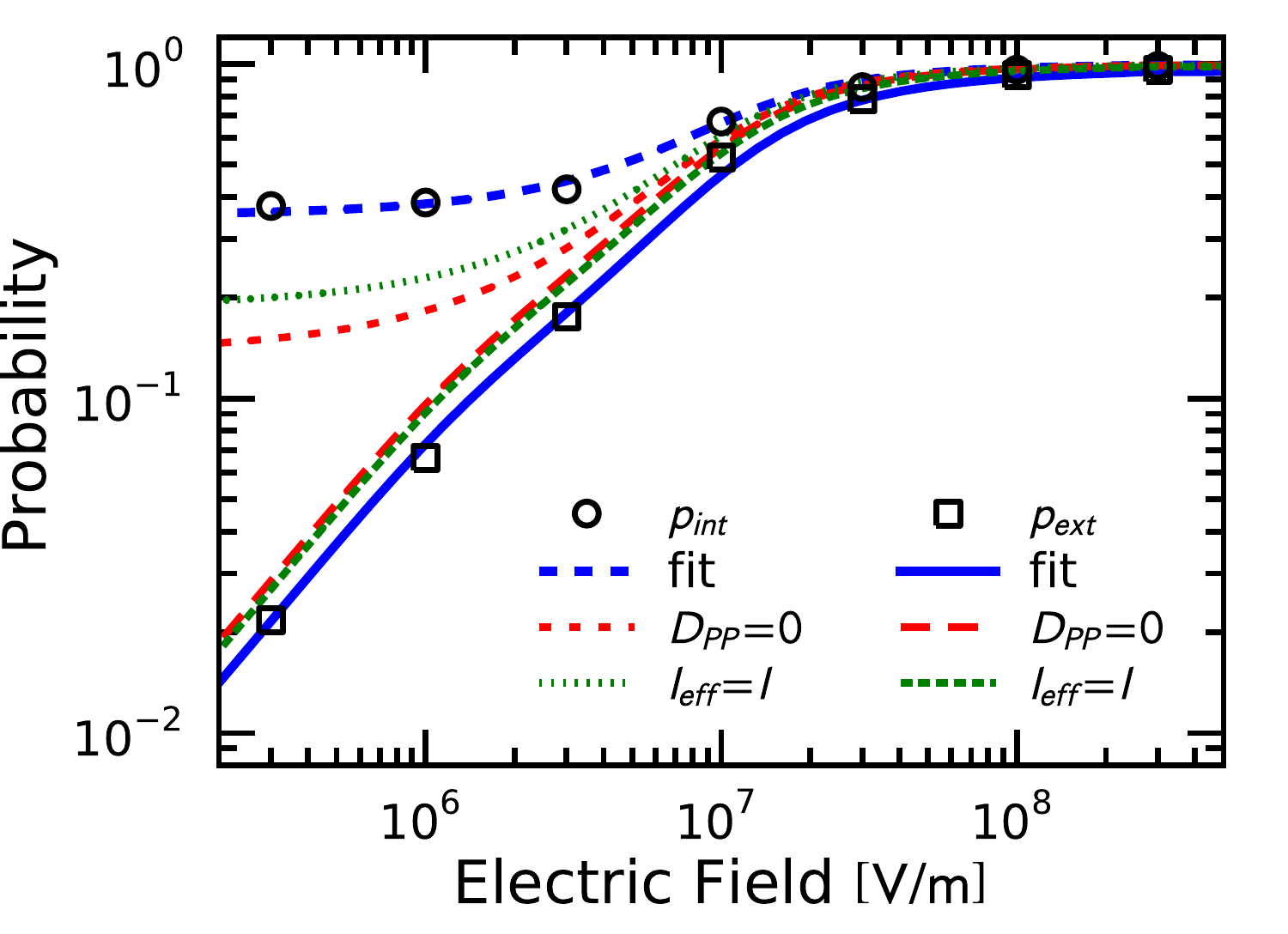}
	\caption{(Color Online) Probability for net photocurrent $p_{ext}$ and PP dissociation $p_{int}$ from Monte Carlo simulation (markers; $1:1$ donor--acceptor blend, CL$=6$, $\tau_{eff}=100$~ns, $300$~K, $l=100$~nm and $\epsilon=3.0$) and analytic calculations (fit; $D_{PP} = 3.8 \cdot 10^{-10}$~m$^2$/s, $l_{eff} = 22$~nm, $r_{PP} = 3.0$~nm, $\mu= 4.3 \cdot 10^{-8}$~m$^2$/Vs , $\tau_{eff}=100$~ns, $300$~K, $l=100$~nm and $\epsilon=3.0$). For comparison, calculations without PP diffusion ($D_{PP}=0$~m$^2$/s) and without virtual device shortening ($l_{eff} = 100$~nm) are shown.\label{fig:4}}
\end{figure}

The application of this new theoretical approach to explain our simulation results is shown in \figref{4}.
Wherever possible, the parameters for the calculations were taken equivalent to those in our simulation, in particular $\tau \equiv \tau_{eff} = 10^{-7}$~s, $T = 300$~K and $l= 100$~nm.
$D_{PP}$ was also set to a value determined in our simulation, i.e., to $D_{PP} = 3.8 \cdot 10^{-10}$~m$^2$/s as average of the two diffusion processes (\figref{2}).
$r_{PP}$, $\mu$ and $l_{eff}$ were taken as fitting parameters.
Best agreement was found for $r_{PP} = 3.0$~nm and $\mu= 4.4 \cdot 10^{-8}$~m$^2$/Vs, values that are in good agreement with our recent publication \cite{deibel2009a}. For $l_{eff} = 22$~nm the device length of $100$~nm is shortened virtually by a factor of $4.5$ for CL$=6$.
We find perfect agreement between our simulation results and the extended theory (\figref{4}).

For comparison, calculations were also done for $D_{PP} = 0$~m$^2$/s and for $l_{eff} = 100$~nm. Clearly, the simulation results cannot be reproduced with these parameter sets.


In conclusion, by using mesoscopic Monte Carlo simulations of polaron pair dissociation and charge extraction in polymer--fullerene blends, we found that the higher local mobility directly increases the polaron pair diffusion length, and delocalisation leads to less hops being needed to overcome a given device length. Thus, the probability of neutral Coulomb bound polaron pairs to accidentally diffuse to an electrode within their given lifetime grows with increasing delocalisation length. In the framework of our Monte Carlo simulation, these effects lead to a high loss of polaron pairs at the surface, thereby reducing the photocurrent. 

In order to describe our findings analytically, we presented a unified model consisting of exact Onsager theory for polaron pair dissociation, the Sokel--Hughes model for charge extraction, and our extension to account for losses by polaron pair diffusion. Applied to organic semiconductor devices, this extended model leads to a deeper insight into the relation of local morphology and the photocurrent.  This has important consequences for the optimization of organic optoelectronic and photovoltaic devices,   highlighting the need to reduce surface losses by adjusting the device configuration in view of blocking layers or selective electrodes. Indeed, in bulk heterojunction solar cells with conjugated polymer donors inhibiting high conjugation lengths, blocking layers not only suppress the extraction of free polarons at the wrong electrodes. They also minimize the diffusion of Coulomb bound polaron pairs to the electrodes, thus increasing their dissociation and extraction probability.


\begin{acknowledgments}

T.S. thanks Mariusz Wojcik for sharing his program to calculate exact Onsager theory. \cite{wojcik2009}
C.D. gratefully acknowledges the support of the Bavarian Academy of Sciences and Humanities. 
The current work is supported by the Deutsche Forschungsgemeinschaft in the framework of the SPP1355 project Phorce and by the Bavarian Ministry of Economic Affairs, Infrastructure, Transport and Technology.

\end{acknowledgments}

\end{document}